# The substantive and practical significance of citation impact differences between institutions: Guidelines for the analysis of percentiles using effect sizes and confidence intervals

Richard Williams** and Lutz Bornmann*

**Department of Sociology

810 Flanner Hall

University of Notre Dame

Notre Dame, IN 46556 USA

E-mail: Richard.A.Williams.5@ND.Edu

Web Page: http://www.nd.edu/~rwilliam/

*Division for Science and Innovation Studies

Administrative Headquarters of the Max Planck Society

Hofgartenstr. 8,

80539 Munich, Germany.

E-mail: bornmann@gv.mpg.de

**Abstract**

In our chapter we address the statistical analysis of percentiles: How should the citation impact of institutions be compared? In educational and psychological testing, percentiles are already used widely as a standard to evaluate an individual's test scores – intelligence tests for example – by comparing them with the percentiles of a calibrated sample. Percentiles, or percentile rank classes, are also a very suitable method for bibliometrics to normalize citations of publications in terms of the subject category and the publication year and, unlike the mean-based indicators (the relative citation rates), percentiles are scarcely affected by skewed distributions of citations. The percentile of a certain publication provides information about the citation impact this publication has achieved in comparison to other similar publications in the same subject category and publication year. Analyses of percentiles, however, have not always been presented in the most effective and meaningful way. New APA guidelines (American Psychological Association, 2010) suggest a lesser emphasis on significance tests and a greater emphasis on the substantive and practical significance of findings. Drawing on work by Cumming (2012) we show how examinations of effect sizes (e.g. Cohen's d statistic) and confidence intervals can lead to a clear understanding of citation impact differences.

Introduction

Researchers in many fields, including bibliometricians, have typically focused on the statistical significance of results. Often, however, relatively little attention has been paid to substantive significance. Chuck Huber (2013, p. 1) provides an example of the possible fallacies of the typical approaches:

> What if I told you that I had developed a new weight-loss pill and that the difference between the average weight loss for people who took the pill and those who took a placebo was statistically significant? Would you buy my new pill? If you were overweight, you might reply, "Of course!" … Now let me add that the average difference in weight loss was only one pound over the year. Still interested? My results may be statistically significant but they are not practically significant. Or what if I told you that the difference in weight loss was not statistically significant — the p-value was "only" 0.06 — but the average difference over the year was 20 pounds? You might very well be interested in that pill. The size of the effect tells us about the practical significance. P-values do not assess practical significance.

The American Psychological Association (APA) (2010) has recently called on researchers to pay greater attention to the practical significance of their findings. Geoff Cumming (2012) has taken up that challenge in his book, *Understanding the New Statistics: Effect Sizes, Confidence Intervals, and Meta-Analysis*. The need for the methods he outlines is clear: despite the serious flaws of approaches that only examine statistical significance, Tressoldi, Giofre, Sella, and Cumming (2013) find that "Null Hypothesis Significance Testing without any use of confidence intervals, effect size, prospective power and model estimation, is the prevalent statistical practice used in articles published in *Nature*, 89%, followed by articles published in *Science*, 42%. By contrast, in all other journals [*The New England Journal of Medicine*, *The Lancet*, *Neuropsychology*, *Journal of Experimental Psychology-Applied*, and the *American Journal of Public Health*], both with high and lower impact factors, most articles report confidence intervals and/or effect size measures." In bibliometrics, it has been also recommended to go beyond statistical significance testing (Bornmann & Leydesdorff, 2013; Schneider, 2012).

In this chapter we review some of the key methods outlined by Cumming (2012), and show how they can contribute to a meaningful statistical analysis of percentiles. The percentile of a certain publication provides information about the citation impact this publication has achieved in comparison to other similar publications in the same subject category and publication year. Following Cumming's (2012) lead, we explain what effect sizes and confidence intervals (CIs) are. We further explain how to assess the ways in which the percentile scores of individual institutions differ from some predicted values and from each other; and how the proportions of highly cited papers (i.e. the top 10% most frequently cited papers) can be compared across institutions. Throughout, our emphasis will be in not only demonstrating whether or not statistically significant effects exist, but in assessing whether the effects are large enough to be of practical significance.

We begin by discussing the types of measures that bibliometricians will likely wish to focus on when doing their research. Specifically, we argue that percentile rankings for all papers, and the proportion of papers that are among the top 10% most frequently cited, deserve special consideration.

Percentile rankings

Percentiles used in bibliometrics provide information about the citation impact of a focal paper compared with other comparable papers in a reference set (all papers in the same research field and publication year). For normalizing a paper under study, its citation impact is evaluated by its rank in the citation distribution of similar papers in the corresponding reference set (Leydesdorff & Bornmann, 2011; Pudovkin & Garfield, 2009). For example, if a paper in question was published in 2009 and was categorized by Thomson Reuters into the subject category "physics, condensed matter", all papers published in the same year and subject category build up its reference set. Using the citation ranks of all papers in the reference set, percentiles are calculated which also lead to a corresponding percentile for the paper in question. This percentile expresses the paper's citation impact position relative to comparable papers.

This percentile-based approach arose from a debate in which it was argued that frequently used citation impact indicators based on using arithmetic averages for the normalization – e.g., "relative citation rates" (Glänzel, Thijs, Schubert, & Debackere, 2009; Schubert & Braun, 1986) and "crown indicators"(Moed, De Bruin, & Van Leeuwen, 1995; van Raan, van Leeuwen, Visser, van Eck, & Waltman, 2010) – had been both technically (Lundberg, 2007; Opthof & Leydesdorff, 2010) and conceptually (Bornmann & Mutz, 2011) flawed. Among their many advantages, percentile rankings limit the influence of extreme outliers. Otherwise, a few papers with an extremely large number of citations could have an immense impact on the test statistics and parameter estimates.

An example will help to illustrate this. The Leiden Ranking uses citation impact indicators based on using arithmetic averages for the normalization (the mean normalized citation score, MNCS) and based on percentiles ($PP_{top\ 10\%}$). For the University of Göttingen, an extreme outlier leads to a large ranking position difference between MNCS and $PP_{top\ 10\%}$:

> "This university is ranked 2nd based on the MNCS indicator, while it is ranked 238th based on the $PP_{top\ 10\%}$ indicator. The MNCS indicator for University of Göttingen turns out to have been strongly influenced by a single extremely highly cited publication. This publication … was published in January 2008 and had been cited over 16,000 times by the end of 2010. Without this single publication, the MNCS indicator for University of Göttingen would have been equal to 1.09 instead of 2.04, and University of Göttingen would have been ranked 219th instead of 2nd. Unlike the MNCS indicator, the $PP_{top\ 10\%}$ indicator is hardly influenced by a single very highly cited publication. This is because the $PP_{top\ 10\%}$ indicator only takes into account whether a publication belongs to the top 10% of its field or not. The indicator is insensitive to the exact number of citations of a publication (Waltman et al, 2012, p. 2425).

Since the percentile approach has been acknowledged in bibliometrics as a valuable alternative to the normalization of citation counts based on mean citation rates, some different percentile-based approaches have been developed (see an overview in Bornmann, Leydesdorff, & Mutz, 2013). More recently, one of these approaches (named $PP_{top\ 10\%}$, also known as the Excellence Rate, which measures the proportion of papers among the 10% most frequently cited papers in a subject category and publication year) has been prominently used in the Leiden Ranking (Waltman et al., 2012) and the SCImago institutions ranking (SCImago Reseach Group, 2012) as evaluation tools.

Three steps are needed in order to calculate percentiles for a reference set: First, all papers in the set are ranked in ascending order of their numbers of citations. Second, each paper is assigned a percentile based on its rank (percentile rank). Percentiles can be calculated in different ways (Bornmann, et al., 2013; Cox, 2005; Hyndman & Fan, 1996). The most commonly used formula is $(100*(i-1)/n)$, where n is the total number of papers, and i the rank number in ascending order. For example, the median value or the 50th percentile rank separates the top-half of the papers from the lower half. However, one can also calculate percentiles as $(100*(i/n))$. This calculation is used, for example, by InCites (Thomson

Reuters, see below). Third, the minimum or maximum of the percentile rank can be adjusted. Papers with zero citations can be assigned a rank of zero. By assigning the rank zero to the papers with zero citations, one ensures that the missing citation impact of these papers is reflected in the percentiles in the same way in every case. Different ranks for papers with zero citations would arise if percentiles are calculated without using a constant rank of zero at the bottom (Leydesdorff & Bornmann, 2012; Zhou & Zhong, 2012).

A technical issue in the case of using percentiles for research evaluation pertains to the handling of ties (e.g., Pudovkin & Garfield, 2009; Rousseau, 2012; Schreiber, 2013, in press; Waltman & Schreiber, 2013). Imagine 50 papers with 61,61,61,58, 58, 58, 58, 58, 58, 58 citations, rest (40 papers) = 1 citations. For this fictitious reference set it is not possible to calculate exactly the top 10% most frequently cited papers. You can take 3/50 (6%) or 10/50 (20%). Thus, the tying of the ranks at the threshold level generates an uncertainty (Leydesdorff, 2012). Schreiber (2012) and Waltman and Schreiber (2013) solved this problem by proposing fractional counting in order to attribute the set under study to percentile rank classes that are pre-defined (for example, the proportion of the top 10% most frequently cited papers, $PP_{top10\%}$).

By proportional attribution of the fractions to the different sides of the threshold, the uncertainty can be removed from the resulting indicator. However, this approach can only be used to determine the exact proportion of $PP_{top\ x\%}$ (e.g. 10) papers in a reference set, but cannot be used for the calculation of percentile ranks of the individual papers under study. Furthermore, the fractional attribution of percentile ranks is computationally intensive. Since individual papers are the units of analysis in many studies, the fractional attribution of percentile ranks is not functional in many situations.

## Data and statistical software

Publications produced by three research institutions in German-speaking countries from 2001 and 2002 are used as data. Institutions 1, 2, and 3 have 268, 549, and 488 publications respectively, for 1305 publications altogether. The percentiles for the publications were obtained from InCites (Thomson Reuters, http://incites.thomsonreuters.com/), which is a web-based research evaluation tool allowing the assessment of the productivity and citation impact of institutions. Percentiles are defined by Thomson Reuters as follows: The percentile in which the paper ranks in its category and database year, based on total citations received by the paper. The higher the number of citations, the smaller the percentile number. The maximum percentile value is 100, indicating 0 cites received. Only article types *article, note, and review* are used to determine the percentile distribution, and only those same article types receive a percentile value. If a journal is classified into more than one subject area, the percentile is based on the subject area in which the paper performs best, i.e. the lowest value (see http://incites.isiknowledge.com/common/help/h_glossary.html). Since in a departure from convention low percentile values mean high citation impact (and vice versa), the percentiles received from InCites are inverted percentiles. To identify papers which belong to the 10% most frequently cited papers within their subject category and publication year ($PP_{top\ 10\%}$), publications from the universities with an inverted percentile smaller than or equal to 10 are coded as 1; publications with an inverted percentile greater than 10 are coded as 0.

For the calculation of the statistical procedures, we used Stata (StataCorp., 2013). However, many other statistical packages could also be used for these calculations (e.g. SAS or R).

Effect Sizes and related concepts

Cumming (2012, p. 34) defines an effect size as the amount of something that might be of interest. He offers several examples. In Table 1, we present measures of effect size that we think are of special interest to bibliometricians.

Table 1: Examples of effect size measures for bibliometric analyses

| Sample Effect Size | Example |
| --- | --- |
| Mean (M) | Mean percentile ranking, e.g. institution's average percentile ranking is 20 |
| Difference between two means | Institution A's average percentile ranking is 40, Institution B's is 50, for a 10 percentage point difference |
| Cohen's d (both for individual institutions and for institutional comparisons) | The average effect of institution type (A or B) on percentile rankings is .25 |
| Proportion* | 20% of the institution's publications are $PP_{top\ 10\%}$ |
| Relative Proportions and/or differences in proportions* | Institution B is twice as likely to have $PP_{top\ 10\%}$ as is institution A |

* Cumming (2012) uses the terms risk and relative risk. His examples refer to accidents. But we can also think of "risk" as pertaining to other events that might happen, e.g. a published paper is "at risk" of becoming highly cited.

In isolation, however, effect sizes have only limited utility. First, because of sampling variability, estimated effect sizes will often be larger or smaller than the true effect is, i.e. just by chance alone an institution's performance could appear to be better or worse than it truly is; or apparent differences between institutions could seem larger or smaller than they actually are. Second, we need a criterion by which effect sizes can be evaluated. A common criterion is to look at statistical significance, e.g. are the differences between two institutions so large that they are unlikely to be due to chance alone? The APA, however, has called on researchers to go beyond statistical significance and assess substantive significance as well. This can be done both via theory (e.g. theory or past experience might say that a 5 point difference between institutions is substantively important while a 1 point difference is not) and via empirical means (using suggested guidelines for when effects should be considered small, moderate, or large).

Therefore, when discussing effect sizes, we present not only the measure of effect size itself, but the related measures that are needed to assess the statistical and substantive significance of the measure. We begin with the mean percentile ranking.

*The Mean.* Table 2 presents the mean percentile rankings for the three institutions along with related measures.

Table 2. Effect Sizes and Significance tests using Mean Percentile Rankings for Individual Institutions

| Statistical Measure | Institution 1 | Institution 2 | Institution 3 |
|---|---:|---:|---:|
| Mean | 49.67 | 32.15 | 45.98 |
| Standard Deviation | 30.66 | 27.49 | 29.40 |
| Standard Error of the Mean | 1.87 | 1.17 | 1.33 |
| Lower bound of the 95% CI | 45.99 | 29.85 | 43.37 |
| Upper Bound of the 95% CI | 53.36 | 34.46 | 48.59 |
| T (for test of $\mu = 50$) | -0.17 | -15.21 | -3.02 |
| N | 268 | 549 | 488 |
| P value (two—tailed test) | .8613 | .0000 | .003 |
| Cohen's d | -.011 | -.649 | -.137 |

The mean is one of the simplest and most obvious measures. It is simply the arithmetic average of the rankings for all the papers published for an institution. Because of the way percentile ranking is coded (see above), a lower score is a better score. There are obvious ways to use the mean to assess the citation impact of an institution. The population mean (50) is known. We can tell at a glance whether an institution is above the average or below it. However other criteria can also be used. An institution may wish to compare itself with the known values of its peer institutions or aspirational peers. Peer institutions might include other elite universities; schools located in the same general geographic area; or colleges that an institution competes against for students and grant money. Aspirational peers might include schools that are known to currently be stronger but who set a standard of excellence that the school is striving to achieve. Hence a university that considers itself among the best in the country or even the world might feel that its publications should average at least among the top 25%. Conversely a school that sees itself as less research oriented might feel that an average ranking of 75 is sufficient if that is what its regional competitors are achieving.

Table 2 shows us that Institution 1 scores barely better than the mean (49.67), Institution 3 is about 4 points better than average (45.98), while Institution 2 is nearly 18 points better than average (32.15).

However, the mean should NOT be the only measure used to assess citation impact. The mean is merely a point estimate. Chance factors alone could increase or lower the value of the estimated mean. CIs (Confidence Intervals) therefore provide a more detailed way of assessing the importance of mean scores. Cumming (2012) and others discuss several ways in which CIs can be interpreted and used for assessment purposes. CIs provide a feel for the precision of measures. Put another way, they show the range that the true value of the mean may plausibly fall in. For example, if the observed mean was 40, the 95% CI might range between 35 and 45. So, while 40 is our "best guess" as to what the mean truly is, values ranging between 35 and 45 are also plausible alternative values.

CIs also provide an approach to hypothesis testing. If the hypothesized value (e.g. the population mean of 50) falls within the CI, we do not reject the null hypothesis. Put another way, if the hypothesized value of 50 falls within the CI, then 50 is a plausible alternative value for the mean and hence cannot be ruled out as a possibility.[1]

Table 2 shows us the CIs for each of the three institutions, but Figure 1 provides a graphical display of the same information that may be easier to understand. At a glance, we can see what the mean for each institution is and what the range of plausible values for the mean is. The horizontal line for mean = 50 makes it easy to see whether the CI does or does not include the value specified by the null hypothesis (the citation impact is equal to a medium impact, i.e. the population mean of 50). If the horizontal line passes through the CI we do not reject the null hypothesis; otherwise we do.

Figure 1. Average Percentile Score by Institution, with 95% CIs

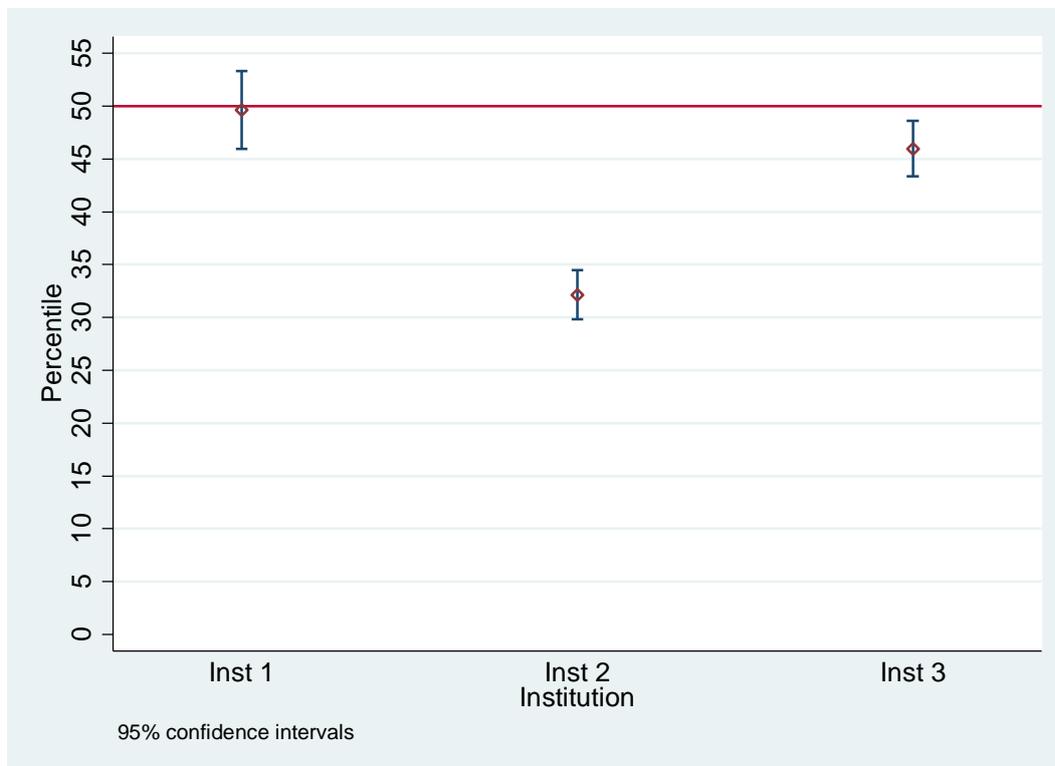

---

[1] Cumming (2012) refers to the CI obtained from an analysis as "One from the dance." What he means is that it is NOT correct to say that there is a 95 percent chance that the true value of the mean lies within the confidence interval. Either the true value falls within the interval or it doesn't. It is correct to say that, if this process were repeated an infinite number of times, then 95 percent of the time the CI would include the true value of the mean while 5 percent of the time it would not. Whether it does in the specific data we are analyzing, we don't know.

For institution 1, the CI ranges between 45.99 (about four points better than average) to 53.36. Because the average value of 50 falls within that interval, we cannot rule out any of the possibilities that Institution 1 is below average, average, or above average. The CI for institution 2 ranges from 29.85 to 34.46, suggesting that it is almost certainly well above average. For institution 3 the CI range is 43.37 to 48.59, implying that it is probably at least a little better than average.

Significance tests (in this case, one-sample t-tests) can also be used to see whether the difference between the observed mean for an institution and the hypothesized mean is statistically significant. In our case we test whether the institutional mean differs from the known population mean of 50, but another criterion could be used if it was thought that a higher or lower criterion was appropriate. The formula for the one-sample t-test, along with the specific calculation for Institution 2, is

$$t = \frac{\bar{x} - \mu_0}{s/\sqrt{N}} = \frac{32.15 - 50}{27.49/\sqrt{549}} = \frac{-17.85}{1.1732} = -15.21$$

where $\bar{x}$ = the sample mean, $\mu_0$ is the value for the mean specified under the null hypothesis (in this case 50), s is the standard deviation of x, and N is the sample size. $s/\sqrt{N}$ is also known as the standard error of the mean. Both the standard deviation of x and the standard error of the mean are reported in table 2.

If the null hypothesis is true – in this case, if the population mean of Institution 2 really is 50 – the t statistic will have a t distribution with N – 1 degrees of freedom. The larger in magnitude the t statistic is (positive or negative), the less likely it is that the null hypothesis is true. The critical value for the t statistic, i.e. the value at which we conclude that the null hypothesis is unlikely to be true, depends on the sample size. In samples this large, the absolute value of the t statistic needs to be 1.96 or greater for us to conclude that observed deviations from the null hypothesis are probably not just due to chance factors alone. The t-test in table 2 shows us that the difference between the mean for Institution 1 and the population mean is not statistically significant. The t values for Institutions 2 and 3 are statistically significant, but the t value for Institution 2 is more than 5 times as large as the t value for Institution 3.

Significance tests have their own limitations though. In particular, statistically significant results are not necessarily substantively meaningful. Significance tests are strongly affected by sample size. If the sample is large enough even trivial differences can be statistically significant. Conversely if the sample is small even a seemingly large difference may not achieve statistical significance. For example, if the sample is large enough a mean score of 49 may statistically differ from the population mean of 50 at the .05 level of significance. Conversely if a sample is small a mean score of 40 might only be statistically significant at, say, the .06 level. As argued earlier, while significance tests can be helpful their utility is also limited.

To make this important point clear, consider an example that is similar to our opening weight loss example: If you were told that an institution had scores that were statistically significantly above average, would you be impressed? Perhaps. But if you were also told that it was 1 point better than average and that this was statistically significant at the .04 level, would you still be impressed? Probably not. Conversely, if you were told that an institution's scores were not statistically significant from the average, would you be unimpressed? Perhaps. But if you were told that its observed score was 10 points better than average and that the difference was statistically significant at the .06 level, would you be impressed then?

Probably in most cases more people would be impressed by the latter institution, even though its higher scores barely missed being statistically significant at the .05 level.

We may have enough of a theoretical or intuitive feel to decide whether an effect is large enough to care about, e.g. theory or intuition or past experience may tell us that a 1 point difference from the mean isn't worth caring about while a 10 point difference is. However, in situations that are less clear, measures such as Cohen's d give us another way of assessing the substantive significance of effects.

### *Cohen's d (for individual institutions)*

Table 2 also includes the Cohen's d statistic for each institution. Cohen's d, and related measures (e.g. Cohen's h, see below), try to illustrate the magnitude of an effect. Put another way, they try to illustrate the substantive importance, as opposed to the statistical significance, of results. Cohen's d and similar measures may be especially useful when it is not otherwise clear whether an effect is large or small. So, for example, if an institution scores 3 points above the mean, should that be considered a substantively large difference, a substantively small difference, or what?

Those with expertise in a well understood field may be able to offer an informed opinion on that. But in less clear situations, Cohen's d provides a potentially superior means of assessing the substantive significance of results as opposed to simply noting what the observed differences are. As Cumming (2012) notes, Cohen's d "can help ES [effect size] communication to a wide range of readers, especially when the original units first used to measure the effect are not widely familiar" (p. 282). Cumming (2012) also notes that Cohen's d can be useful in meta-analysis where different researchers have measured key variables in different ways.

In the single sample case, Cohen's d equals the difference between the observed mean and the hypothesized mean (e.g. 50) divided by the sample standard deviation, i.e.

$$\text{Cohen's d} = \frac{\bar{x} - \mu_0}{s}$$

Through simple algebra, it can also be shown that Cohen's d = $t/\sqrt{N}$. So, for example, the Cohen's d value for institution 2 is

$$\text{Cohen's d} = \frac{\bar{x} - \mu_0}{s} = \frac{32.15 - 50}{27.49} = \frac{-17.85}{27.49} = -.649$$

As Cohen (1988) notes, Cohen's d is similar to a z score transformation. For example, a Cohen's d value of .2 would mean that the mean in the sample was .2 standard deviations higher than the hypothesized mean. Cohen (1988) suggested that effect sizes of 0.2, 0.5, and 0.8 (or, if the coding is reversed, -.2, -.5 and -.8) correspond to small, medium and large effects.

Put another way, the usefulness of Cohen's d depends, in part, on how obvious the meaning is of observed differences. If, for example, we knew that students in an experimental teaching program scored one grade level higher than their counterparts in traditional programs, such a difference might have a great deal of intuitive meaning to us. But if instead we knew that they scored 7 points higher on some standardized test, something like Cohen's d could help us to assess how large such a difference really is. [2]

---

[2] Cumming (2012) notes various cautions about using Cohen's d (p. 283). For example, while it is common to use sample standard deviations as we do here, other "standardizers" are possible, e.g. you might use the standard deviation for a reference population, such as elite institutions. Researchers should be clear exactly how Cohen's d was computed.

Effect Sizes - Williams & Bornmann – April 11, 2014                                        Page 16

Returning to table 2 and our three institutions, the Cohen's d statistic for Institution 1 is, not surprisingly, extremely small, almost 0. For Institution 3, even though the difference between the school's sample mean and the population mean 50 is statistically significant, the Cohen's d statistic is only -.137. This falls below the value of -.2 that Cohen (1988) had suggested represented a small effect. For Institution 2, Cohen's d is equal to -.649. This falls almost exactly halfway between Cohen's suggested values of -.5 for medium and -.8 for large. Therefore, if we had no other clear criteria for assessing the magnitude of effects, Cohen's d would lead us to conclude that differences between institutions 1 and 3 and the population mean of 50 are not substantively important, while the difference between Institution 2 and the population mean is moderately to highly important.

Before leaving Table 2, it should be noted that we did additional analyses to confirm the validity of our results. In both Tables 2 and 3, we make heavy use of t tests and related statistics. As Acock (2010) points out, t tests assume that variables are normally distributed; and, when two groups are being compared, it is often assumed that the variances of the two groups are equal. Percentile rankings violate these assumptions in that they have a uniform, rather than normal, distribution. Luckily, Acock adds that t tests are remarkably robust against violations of assumptions.

Nonetheless, to reassure ourselves that our results are valid, we double-checked our findings by using techniques that are known to work well when distributional assumptions are violated. In particular, for both Tables 2 and 3, we verified our findings using bootstrapping techniques. Bootstrapping is often used as an alternative to inference based on parametric assumptions when those assumptions are in doubt (Cameron & Trivedi, 2010). Bootstrapping resamples observations (with replacement) multiple times. Standard errors, confidence intervals and significance tests can then be estimated from the multiple resamples. Bootstrapping produced significance tests and confidence intervals that were virtually identical to those reported in our tables, giving us confidence that our procedures are valid.

*Mean differences between institutions*

Rather than use some absolute standards for assessment (e.g. is the average score for an institution above or below the population average?) we may wish to compare institutions against each other. For example, a school might wish to compare itself against a school that it considers its rival, or that competes for students in the same geographic area. Does one institution have average scores that are significantly higher than the other's, or are their scores about the same? Alternatively, we might want to compare the same institution at two different points in time – have its average scores gotten better across time or have they gotten worse? Table 3 presents such comparisons for the institutions in our study.

Table 3. Effect Sizes and Significance tests for Differences in Percentile Rankings Across Institutions

| Statistical Measure | Institution 1 Vs Institution 2 | Institution 1 Vs Institution 3 | Institution 3 Vs Institution 2 |
|---|---|---|---|
| Difference between Means | 17.52 | 3.69 | 13.83 |
| Standard Deviation (pooled) | 28.57 | 29.85 | 28.40 |
| Standard Error of the Mean Difference | 2.13 | 2.27 | 1.77 |
| Lower bound of the 95% CI for the difference | 13.34 | -.76 | 10.36 |
| Upper Bound of the 95% CI for the difference | 21.70 | 8.15 | 17.30 |
| T (for test of μs are equal) | 8.23 | 1.63 | 7.83 |
| P value (two—tailed test) | .0000 | .1042 | .0000 |
| Cohen's d | .613 | .124 | .487 |

As Table 3 shows, Institution 1 scores 17.52 points worse than Institution 2. Similarly Institution 2 averages almost 14 points better than Institution 3. The mean difference between Institutions 1 and 3 is a much more modest 3.69 points.

Again, simply comparing the means for two institutions is not adequate. Apparent differences may not be statistically significant; just by chance alone one institution could have scored higher than the other. And even if more than chance was likely involved in the differences, the substantive significance of differences still needs to be assessed.

CIs can again be useful. Referring back to figure 1, we can see whether the 95% CIs for two institutions overlap. As Cumming (2012) notes, if they do, then the difference between the institutions is not statistically significant at the .01 level. A common error is to assume that if two 95% CIs overlap then the difference in values is not statistically significant at the .05 level. This is wrong because it is unlikely that, by chance alone, one variable would have an atypically low observed value while the other would have a value that was atypically high.

Even more useful is that we can compute the CI for the difference between the scores of two institutions. If 0 falls within the 95% CI, then the difference between the two groups is not statistically significant. Or, if the observed difference is 10 but the CI ranges between 5 and 15, then the actual difference could plausibly be as low as 5 points or as much as 15. Figure 2 provides a graphical and possibly clearer illustration of the information that is also contained in Table 3. If the horizontal line at y = 0 crosses the CI, we know that the difference between the two means is not statistically significant. The CIs show that the differences between 1 and 3 are modest or even nonexistent while the differences between 2 and the other institutions are large (10 points or more) even at the lower bounds of the CIs.

Figure 2. Differences in Mean Percentile Rankings with 95% CIs

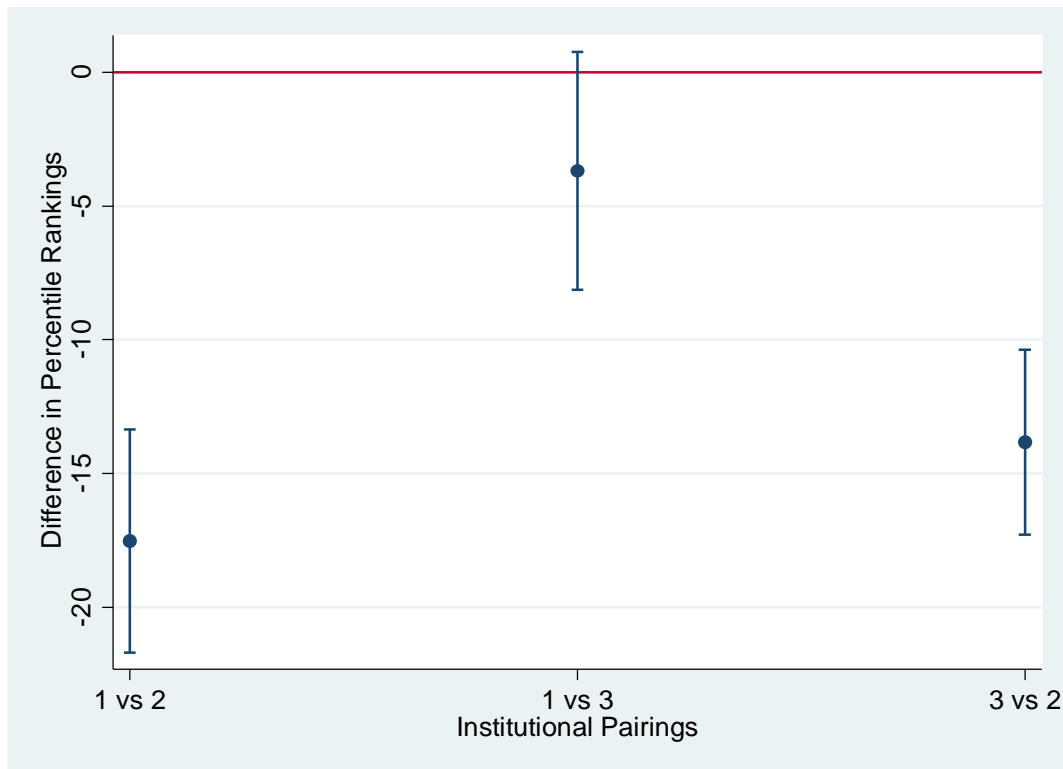

Significance tests (in this case an independent sample t-test) can again be used. Because two groups are being compared, the calculations are somewhat more complicated but still straightforward. It is often assumed that the two groups have the same variance[3]. But, in the samples there are separate estimates of the variance for each group. A pooled estimate for the variance of the two groups is therefore estimated as follows (again we show the general formula, and the specific calculation for institutions 1 and 2).

$$s_p = \sqrt{\frac{(n_1-1)s_1^2 + (n_2-1)s_2^2}{n_1+n_2-2}} = \sqrt{\frac{(268-1)30.66^2 + (549-1)27.49^2}{815}} = \sqrt{816.098} = 28.57$$

---

[3] With independent samples there are two different types of t-tests that can be conducted. The first type, used here, assumes that the variances for each group are equal. The second approach allows the variances for the two groups to be different. In our examples, it makes little difference which approach is used, since, as table 2 shows, the standard deviations for the three groups are similar. In cases where the variances do clearly differ the second approach should be used. Most, perhaps all, statistical software packages can compute either type of t-test easily.

The standard error of the difference (again, both in general, and specifically for institutions 1 and 2) is

$$se_p = \sqrt{s_p^2\left(\frac{N_1+N_2}{N_1 N_2}\right)} = \sqrt{28.57^2\left(\frac{249+568}{249*568}\right)} = \sqrt{28.57^2\left(\frac{817}{141432}\right)} = 2.13$$

The t test (both in general and for institutions 1 and 2) is

$$t = \frac{\overline{x_1} - \overline{x_2}}{se_p} = \frac{49.67 - 32.15}{2.13} = \frac{17.52}{2.13} = 8.23$$

The tests confirm that Institution 2 does better than the other two institutions and that these differences are statistically significant, while the differences between 1 and 3 are small enough that they could just be due to chance factors.

As noted earlier, bootstrapping techniques, which are often used when the validity of parametric assumptions is in doubt (e.g. variables are normally distributed), produced results virtually identical to those reported in Table 3. As an additional check, for Table 3 we also computed Mann-Whitney tests. Mann-Whitney tests are appropriate when dependent variables have ordinal rather than interval measurement (Acock 2010), and percentile rankings clearly have at least ordinal measurement. The Mann-Whitney test statistics were virtually identical to the t test statistics we reported in the table, again increasing our confidence that our results are valid. Further, we think that the approach we are using for Table 3 is superior to nonparametric alternatives such as Mann-Whitney because statistics such as Cohen's d and confidence intervals can be estimated and interpreted, making the substantive significance of results clearer.

Significance tests (t test or Mann-Whitney test) have similar problems as before. If sample sizes are large, even small differences between the two groups can be statistically significant, e.g. a difference of only 1 point could be statistically significant if the samples are large enough. Conversely, even much larger differences (e.g. 10 points) may fail to achieve significance at the .05 level if the samples are small.

To better assess substantive significance, Cohen's d can be calculated for the difference between means. The formula (both in general and for institutions 1 and 2 is)

$$\text{Cohen's d} = \frac{\bar{x}_1 - \bar{x}_2}{s_p} = \frac{49.67 - 32.15}{28.57} = \frac{17.52}{28.57} = .613$$

The Cohen's d indicates that the differences between Institution 2 and Institution 1 (.613) and between Institution 2 and Institution 3 (.487) are at least moderately large. Conversely, the Cohen's d statistic of .124 for comparing institutions 1 and 3 falls below Cohen's suggested level for a small effect.

*Proportions (both for one institution and for comparisons across institutions).*
As noted above, one way of evaluating institutions is to see how their average scores compare. However, it could be argued that evaluations should be made, not on average scores, but on how well an institution's most successful publications do. In particular, what proportion of an institution's publications rank among the 10% most frequently cited papers?

Again, there is an obvious criterion: overall we know that 10% of all papers rank among the top 10% of those most cited. We use that criterion here, but the criterion could be made higher or lower as deemed appropriate for the type of institution.

There are important differences in how statistics and significance tests are computed for binary outcomes. Binary variables do not have a normal distribution, nor are their means and variances independent of each other. If the mean of Y = P, then V(Y) = P(1-P), e.g. if there is a .3 probability that Y = 1, then V(Y) = .3*7 = .21. As the Stata 13 Reference Manual (2013) points out, several different formulas have been proposed for computing confidence intervals (e.g. the Wilson, Agresti, Klopper-Pearson, and Jeffries methods) and several other statistics. We use the large sample methods used by Stata's prtest command, but researchers, especially those with smaller samples, may wish to explore other options.

But, other than that, the arguments are largely the same as earlier. Significance tests and CIs have the same strengths and weaknesses as before. Visual means of presenting results are also quite similar. Effect size measures help to provide indicators of the substantive significance of findings. In short, the most critical difference from before is that a different criterion is being used for the assessment of impact.

Table 4 presents the effect sizes and related measures for $PP_{top\ 10\%}$ for each institution separately. Figure 3 illustrates how the proportions and their CIs can be graphically depicted.

Table 4. Effect Sizes and Significance tests for $PP_{top\ 10\%}$ – Individual Institutions

| Statistical Measure | Institution 1 | Institution 2 | Institution 3 |
|---|---:|---:|---:|
| $PP_{top\ 10\%}$* | 11.19 | 29.14 | 11.68 |
| Standard Error* | 1.93 | 1.94 | 1.45 |
| Lower bound of the 95% CI* | 7.42 | 25.34 | 8.83 |
| Upper Bound of the 95% CI* | 14.97 | 32.95 | 14.53 |
| Z (for test of $PP_{top\ 10\%}$ = .10) | .65 | 14.95 | 1.24 |
| P value (two-tailed test) | .51 | .0000 | .22 |
| Cohen's h | .039 | .497 | .054 |
| N | 268 | 549 | 488 |

* Numbers are multiplied by 100 to convert them into percentages

Note that Z tests rather than t tests are used for binomial variables. Further, the observed sample standard deviation is not used in the calculation of the test statistic; rather, the standard deviation implied by the null hypothesis is used. The Z statistic for Institution 2 is calculated as follows:

$$z = \frac{p - p_0}{\sqrt{p_0(1-p_0)/N}} = \frac{.2914 - .10}{\sqrt{.1*.9/549}} = \frac{.1914}{.0128} = 14.95$$

Figure 3. PP$_{top\ 10\%}$ by Institution, with 95% CIs

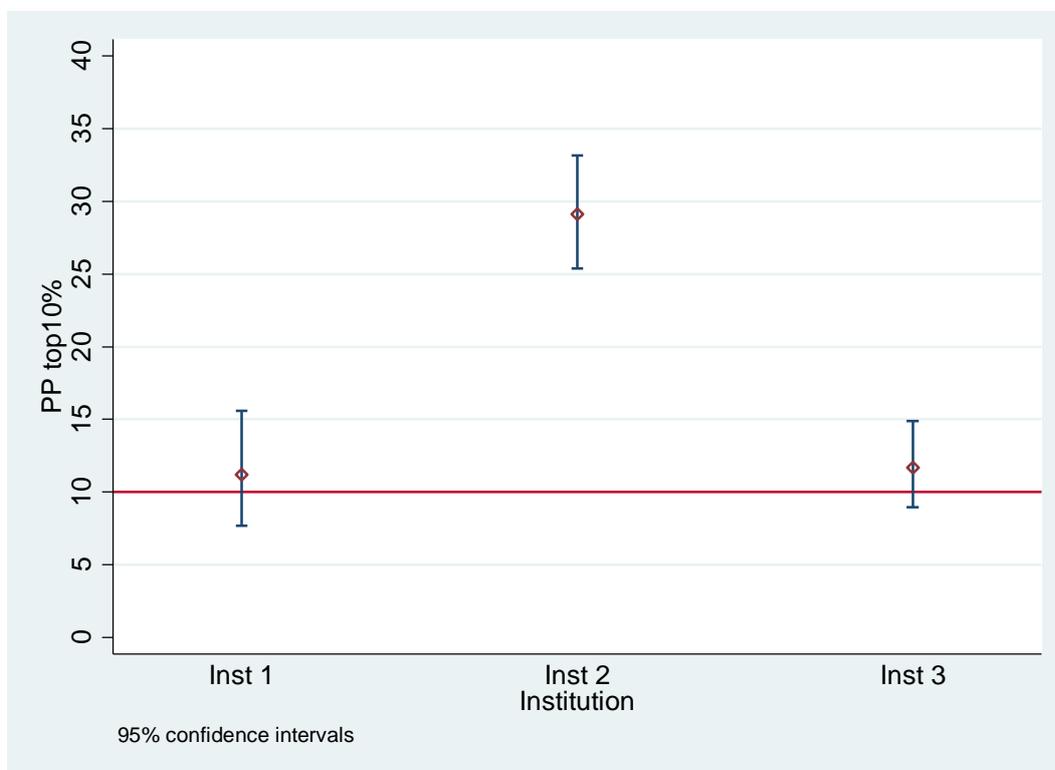

The results are very consistent with what we saw when we analyzed mean percentile rankings. Institutions 1 and 3 are slightly above average in that a little over 11% of their papers rank in the P$_{top\ 10\%}$. However, the CIs for each include 10, and the significance tests also indicate that the null hypothesis that PP$_{top\ 10\%}$ = 10 cannot be rejected. Institution 2, on the other hand, has more than 29% PP$_{top\ 10\%}$. Both the CIs and the significance test indicate that such a strong performance is highly unlikely to be due to chance alone.

Note that the table does not include Cohen's d, because it is not appropriate for dichotomous dependent variables. Instead, for binary variables Cohen (1998) proposes an effect size measure he calls h.[4] The formula is not particularly intuitive, but it has several desirable properties. h is calculated as follows:

$$o = 2*\arcsin\left(\sqrt{P}\right),$$
$$o_0 = 2*\arcsin\left(\sqrt{P_0}\right),$$
$$h = o - o_0$$

So, for example, for Institution 2, P = .2914 (PP$_{top\ 10\%}$=29.14). Since we are using $P_0$ = .10, The h value for institution 2 is

$$o = 2*\arcsin\left(\sqrt{P}\right) = 2*\arcsin\left(\sqrt{.2914}\right) = 1.1404341,$$
$$o_0 = 2*\arcsin\left(\sqrt{P_0}\right) = 2*\arcsin\left(\sqrt{.10}\right) = .64350111,$$
$$h = (o - o_0) = (1.1404341 - .64350111) = .497$$

According to Cohen (1988) the suggested small, medium and large values of .2, .5, and .8, respectively, continue to be reasonable choices for the h statistic, at least when there is little guidance as to what constitutes a small, medium or large effect. He further notes that in the fairly common case of $P_0$ = .5, h will equal approximately 2 * (P - .5).

Table 5 presents the corresponding measures for institutional comparisons.

---

[4] Nonetheless, as we found for other measures in our analysis, Cohen's d seems robust to violations of its assumptions. When we estimated Cohen's d using binary dependent variables, we got almost exactly the same numbers as we did for Cohen's h.

Table 5. Effect Sizes and Significance tests for Differences in $PP_{top\ 10\%}$ Across Institutions

| Statistical Measure | Institution 1 Vs Institution 2 | Institution 1 Vs Institution 3 | Institution 3 Vs Institution 2 |
|---|---:|---:|---:|
| Difference between proportions | -17.95 | -0.49 | -17.47 |
| Standard Error | 2.73 | 2.43 | 2.42 |
| Lower bound of the 95% CI for the difference | -23.31 | -5.22 | -22.21 |
| Upper Bound of the 95% CI for the difference | -12.59 | 4.24 | -12.71 |
| Z (for test of $PP_{top10\%}$ are equal) | -5.70 | -0.20 | -6.90 |
| Cohen's h | -.458 | -.015 | -.443 |
| P value (two—tailed test) | .0000 | .8411 | .0000 |

Again, the results are very consistent with before. The differences between institutions 1 and 3 are very slight and may be due to chance factors alone. Institution 2, on the other hand, has more than twice as many $PP_{top\ 10\%}$ as do institutions 1 and 3, and the differences are highly statistically significant. The calculation of Cohen's h is similar to before, except that $P_2$ is substituted for $P_0$, e.g. for institutions 1 and 2

$$o_1 = 2*\arcsin\left(\sqrt{P_1}\right) = 2*\arcsin\left(\sqrt{.1119}\right) = .68218016,$$
$$o_2 = 2*\arcsin\left(\sqrt{P_2}\right) = 2*\arcsin\left(\sqrt{.2914}\right) = 1.1404341,$$
$$h = (o_1 - o_2) = (.68218016 - 1.1404341) = -.458$$

Incidentally, several other measures of effect size have been proposed and are widely used for analyses of binary dependent variables. These include risk ratios, odds ratios and marginal effects. For a discussion of some of these methods, see Williams (2012), Bornmann and Williams (2013), Deschacht and Engels (this book) and Long and Freese (2006).

Conclusions

The APA has called on researchers to employ techniques that illustrate both the statistical and the substantive significance of their findings. Similarly, in the statistics paragraph of the *Uniform Requirements for Manuscripts* (URM) of the International Committee of Medical Journal Editors (ICMJE, 2010) it is recommended to "describe statistical methods with enough detail to enable a knowledgeable reader with access to the original data to verify the reported results. When possible, quantify findings and present them with appropriate indicators of measurement error or uncertainty (such as confidence intervals). Avoid relying solely on statistical hypothesis testing, such as P values, which fail to convey important information about effect size."

In this chapter, we have shown that the analysis of effect sizes for both means and proportions are worthwhile, but must be accompanied by criteria with which the statistical and the substantive significance of effect sizes can be assessed. Measures of statistical significance are, in general, well known, but we have shown how they can be applied to bibliometric data. Assessment of substantive significance depends, in part, on theory or empirical means: how large does an effect size need to be in order to be considered important? But, when theory and empirical evidence are unclear, measures such as Cohen's d can provide guidelines for assessing effects. As we have seen, effects that are statistically significant may not have much substantive importance. Conversely there may be situations where effects fail to achieve statistical significance but may nonetheless have a great deal of substantive significance. Using tools presented in this paper and in Cumming's (2012) book, researchers can assess both the statistical and substantive significance of their findings.

For those who would like to replicate our findings or try similar analyses with their own data, the Appendix shows the Stata code for the analyses presented in this chapter and for the additional double-checks we did to verify the validity of our results.

# Appendix: Stata Code Used for These Analyses

```
* Stata code for Williams & Bornmann book chapter on effect sizes.
* Be careful when running this code -- make sure it doesn't
* overwrite existing files or graphs that use the same names.
version 13.1
use "http://www3.nd.edu/~rwilliam/statafiles/rwlbes", clear
gen inst12 = inst if inst!=3
gen inst13 = inst if inst!=2
gen inst23 = inst if inst!=1
gen top10 = perc <= 10
* Limit to 2001 & 2002; this can be changed
keep if py <=2002

* Table 2
* Single group designs - pages 286-287 of Cumming
* For each institution, test whether percentile mu = 50
* Note that negative differences mean better than average performance
forval instnum = 1/3 {
        display
        display "Institution `instnum'"
        ttest perc = 50 if inst==`instnum'
        display
        display "Cohen's d = " r(t) / sqrt(r(N_1))
        * DOUBLE CHECK: Compares above CIs and t-tests with bootstrap
        * Results from the test command should be similar to the t-test
        * significance level
        bootstrap, reps(100): reg perc if inst==`instnum'
        test _cons = 50
}

* Table 3
* Two group designs - Test whether two institutions
* differ from each other on mean percentile rating.
* Starts around p. 155
* Get both the t-tests and the ES stats, e.g. Cohen's d
* Note: you should flip the signs for the 3 vs 2 comparison

foreach iv of varlist inst12 inst13 inst23 {
        display      "perc is dependent, `iv'"

        ttest perc, by(`iv')
        scalar n1 = r(N_1)
        scalar n2 = r(N_2)
        scalar s1 = r(sd_1)
        scalar s2 = r(sd_2)
        display
        display "Pooled sd is " ///
                sqrt(((n1 - 1) * s1^2 + (n2 - 1) * s2^2 ) / (n1 + n2 - 2))
        display
        esize two perc, by(`iv') all
        display
        * DOUBLE CHECKS: Compare Mann-Whitney & bootstrap results with above
        * Mann-Whitney test
        ranksum perc, by(`iv')
        * Bootstrap
        bootstrap, rep(100): reg perc i.`iv'
}

* Table 4
* Proportions in Top 10, pp. 399-402
* Single institution tests
* Numbers in tables are multiplied by 100
forval instnum = 1/3 {
        display
        display "Institution `instnum'"
        prtest top10 = .10 if inst==`instnum'
```

```
        display
        display
        scalar phi1 = 2 * asin(sqrt(r(P_1)))
        scalar phi2 = 2 * asin(sqrt(.10))
        di "h effect size = " phi1 - phi2
        display
}

* Table 5
* Proportions in Top 10 - pairwise comparisons of institutions
* Numbers in table are multiplied by 100
foreach instpair of varlist inst12 inst13 inst23 {
        display
        display "`instpair'"
        prtest top10, by (`instpair')
        display
        scalar phi1 = 2 * asin(sqrt(r(P_1)))
        scalar phi2 = 2 * asin(sqrt(r(P_2)))
        di "h effect size = " phi1 - phi2
        display
        * NOTE: Cohen's d provides very similar results to Cohen's h
        esize two top10, by (`instpair') all
        display
}
* Do graphs with Stata
* NOTE: Additional editing was done with the Stata Graph Editor
* Use ciplot for Univariate graphs

* Figure 1 - Average percentile score by inst with CI
ciplot perc, by(inst) name(fig1, replace)

* Figure 3
* Was edited to multiply by 100
ciplot top10, bin by(inst) name(fig3, replace)

*** Save figures before running figure 2 code

* Figure 2 - Differences in mean percentile rankings
* Use statsby and serrbar for tests of group differences
* Note: Data in memory is overwritten
gen inst32 = inst23 * -1 + 4
tab2 inst32 inst23
statsby _b _se, saving(xb12, replace) : reg perc i.inst12
statsby _b _se, saving(xb13, replace) : reg perc i.inst13
statsby _b _se, saving(xb32, replace) : reg perc i.inst32
clear all
append using xb12 xb13 xb32, gen(pairing)
label define pairing 1 "1 vs 2" 2 "1 vs 3" 3 "3 vs 2"
label values pairing pairing
serrbar _stat_2 _stat_5 pairing, scale(1.96) name(fig2, replace)
```